\documentclass{article}

\usepackage{arxiv}

\usepackage[utf8]{inputenc} 
\usepackage[T1]{fontenc}    
\usepackage{hyperref}       
\usepackage{url}            
\usepackage{booktabs}       
\usepackage{amsfonts}       
\usepackage{nicefrac}       
\usepackage{microtype}      
\usepackage{lipsum}
\usepackage{amsmath}
\usepackage{graphicx}
\graphicspath{ {./images/} }

\title{MEMS-tunable dielectric metasurface lens using thin-film PZT for large displacements at low voltages}

\author{
 Christopher A. Dirdal, Paul C. V. Thrane, Firehun T. Dullo, Jo Gjessing, Anand Summanwar, Jon Tschudi \\
  SINTEF Smart Sensors and Microsystems,\\
  Forskningsveien 1,\\
  0373 Oslo \\
  Norway\\
  \texttt{christopher.dirdal@sintef.no} \\
}

\begin{document}
\maketitle
\begin{abstract}
Tunable focusing is a desired property in a wide range of optical imaging and sensing technologies but has tended to require bulky components which cannot be integrated on-chip and have slow actuation speeds. Recently, integration of metasurfaces into electrostatic MEMS architectures has shown potential to overcome these challenges, but has offered limited out of plane displacement range while requiring large voltages. We demonstrate for the first time a movable metasurface lens actuated by integrated thin-film PZT MEMS, which has the advantage of offering large displacements at low voltages. An out of plane displacement of a metasurface in the range of 7.2 µm is demonstrated under a voltage application of 23V. This is roughly twice the displacement at a quarter of the voltage of state of the art electrostatic out of plane actuation of metasurfaces. Utilizing this tunability we demonstrate a varifocal lens doublet with a focal shift on the order of 250$\mu$m at the wavelength $1.55\mu$m. Thin-film PZT MEMS-metasurfaces is a promising platform for miniaturized varifocal components.
\end{abstract}


\section{Introduction}
The ability to vary the focal length of a lens allows for optical probing or imaging at different spatial depths. Such \emph{varifocal} lenses are therefore important components in a wide variety of optical sensor and imaging technologies. 
Varifocal lenses have however tended to be bulky, slow and power intensive owing to the use of bulky refractive (curved) lenses and stepper motors for their realization. For an increasing number of applications, however, the need for dramatically miniaturized optical components is paramount: e.g. in-vivo medical diagnostics for personalized medicine, drone based sensing and imaging (for e.g. environmental monitoring), and wearable devices (e.g. AR glasses, smart watches, cellular phones).

Planar lens technologies, such as diffractive optical elements (DOEs) and recently metasurfaces, offer a route towards miniaturizing lenses. The added light control in the latter platform opens for further miniaturization by including multiple optical functions in a single metasurface \cite{xu2021multifunctional}. Miniaturization on the \emph{system} level can be achieved by making the optical functions of the metasurfaces actively tunable, towards which a lot of research is currently directed.

\begin{figure}[htbp]
\centering
\fbox{\includegraphics[width=0.97\linewidth]{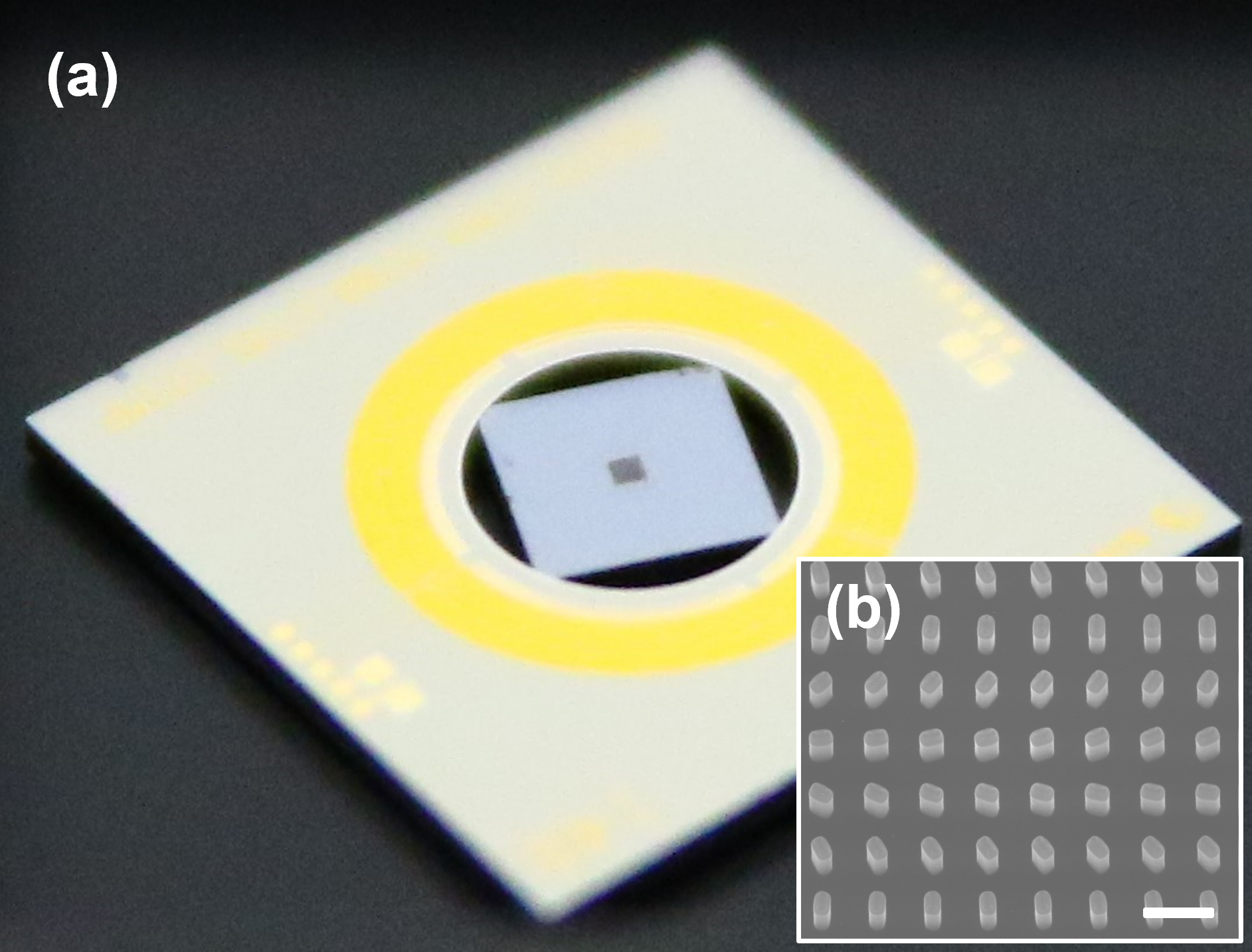}}
\caption{(a) Metasurface on square silicon chip suspended in a thin-film PZT MEMS-actuated ring. (b) SEM image of part of the metasurface structure located at the center of the suspended silicon chip. The scalebar represents 1$\mu$m.}
\label{fig:MEMS_device}
\end{figure}

Tunability of the metasurface structure has for instance been demonstrated through the use of liquid crystals, phase-change materials and two-dimensional materials \cite{kang2019recent}. Another route has been to use mechanical actuation such as stretching elastomeric substrates \cite{kamali2016highly}, and also the use of electrostatic Micro-Electro-Mechanical-Systems (MEMS) actuation \cite{arbabi2018mems, han2020mems, holsteen2019temporal, roy2018dynamic}. The latter has the advantage of fast actuation with the possibility of monolithic integration into standard Silicon MEMS process lines. 

Electrostatic MEMS has recently been used to demonstrate metasurface varifocal components relying on relative displacement of lenses along the optical axis in \cite{arbabi2018mems} and orthogonal to the optical axis in the realization of an Alvarez lens in \cite{han2020mems}. For the former, large focal length tunability requires large out of plane displacement: Using electrostatic actuation, a displacement on the order of 3$\mu$m is achieved by the application of around 80V \cite{arbabi2018mems}. We have recently demonstrated an alternative to electrostatic actuation on metasurfaces: The use of thin-film piezoelectric PZT MEMS actuation for the displacement of a metasurface in which small movements are used to modulate gap surface plasmon resonances \cite{meng2021dynamic}. In this letter we demonstrate that thin-film PZT MEMS can also be used for large range displacement of metasurfaces, relevant for varifocal lens applications at low voltage. Roughly twice the out-of-plane displacement of a dielectric metasurface (on the order of $7.2\mu$m) is achieved compared with that previously reported in \cite{arbabi2018mems} at a quarter of the voltage. We use this capability to demonstrate a focal shift of $250\mu$m in a proof-of-concept demo.



\section{Long displacement range with thin-film PZT}
\label{sec:Displacement}

Achieving effective long range displacement is important for a wide range of micro-optical applications. 
Non-resonant electrostatic MEMS achieves limited displacement range despite requiring large voltages. Piezoelectric thin films, on the other hand, have the advantage of high displacement range at low voltages \cite{kobayashi2007tunable}. Stroke lengths of several tens of micrometers have been demonstrated for piezoelectric cantilevers at voltages below 20V \cite{fang2006modeling}. A drawback of piezo-electric actuation, however, is the presence of hysteresis. Hysteresis can be handled by implementing feedback mechanisms for fine control of the displacement, and the effect of hysteresis can be reduced by poling \cite{bakke2010novel}.

Figure \ref{fig:MEMS_device} displays a thin film PZT MEMS device with a suspended metasurface on a square silicon chip at its center. The visible gold ring is a stack consisting of electrodes placed on top of a PZT membrane and a bottom electrode. Voltage application over the PZT membrane causes it to mechanically deform, moving the center silicon chip. The ring is segmented into two concentric groups of electrodes (three in each ring, six in total). If either all inner or all outer electrodes are actuated, the center chip moves in or out like a piston. The principle of operation is described in greater detail in \cite{bakke2010novel} where a similar device is used to make a micromirror that can be operated in tip-tilt as well as in piston mode. The MEMS is designed to have a switching speed of around 4 kHz when actuating a 3 mm disc of 400 $\mu$m silicon thickness.





\begin{figure}[ht!]
\centering\includegraphics[width=1.0\linewidth]{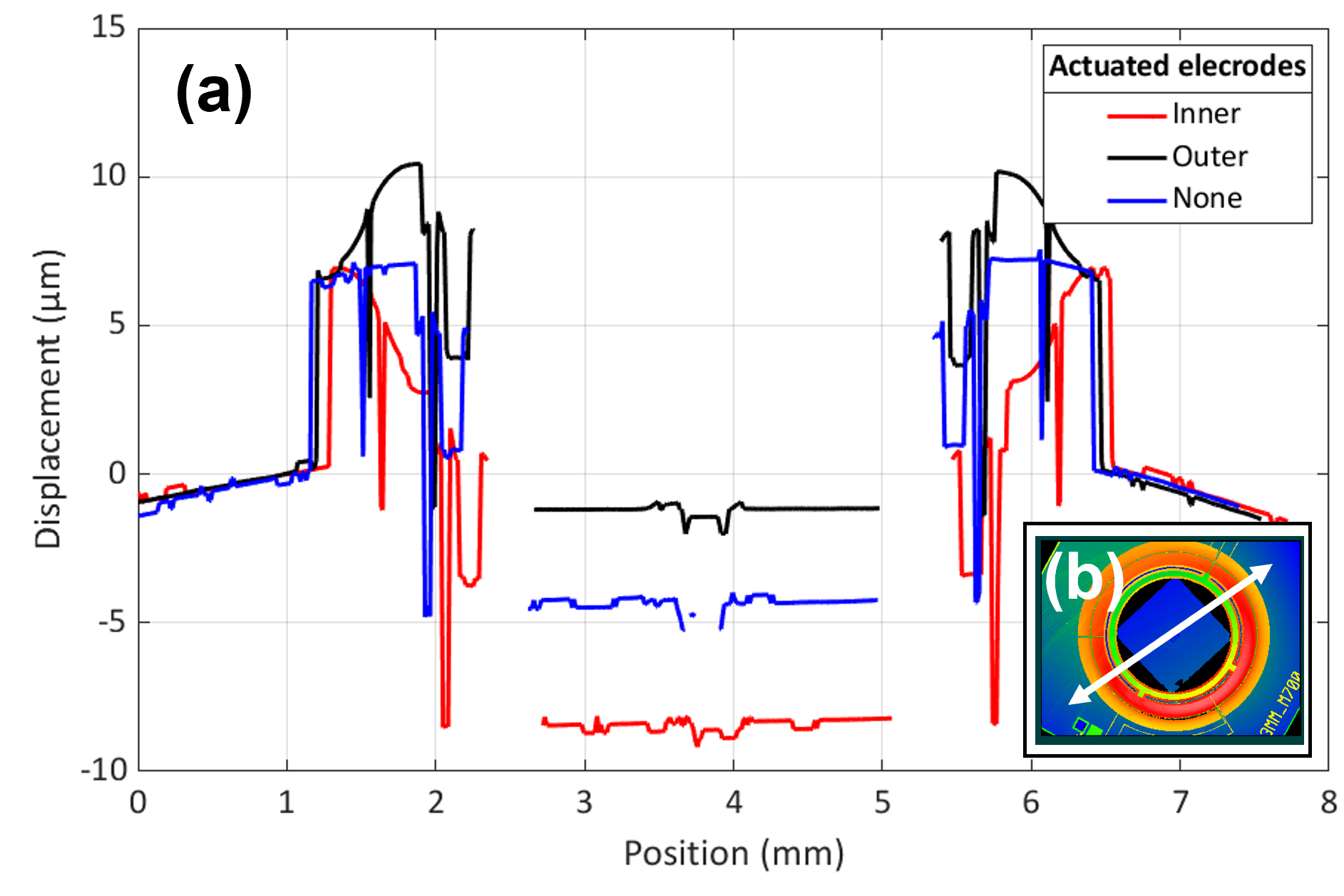}
\caption{(a) Cross-sectional height-profile of MEMS metalens along the line shown in the insert (b) measured by white-light interferometry. The height of the suspended chip with metasurface is changed between three actuation states. It is at its lowest position when 23V is applied to the set of innermost electrodes due to the PZT membrane pulling the ring downwards. The highest position is achieved when 23V is applied to the outermost electrodes, due to the PZT membrane drawing the ring upwards. When no voltage is applied, an intermediate position is attained. A displacement range of 7.2µm is observed, and a high degree of planarity is throughout the movement.}
\label{fig:Zygo}
\end{figure}

In order to measure the displacement range of the thin film PZT MEMS metasurface lens (or \emph{metalens}), a height map was taken for different actuation positions by use of white light interferometry. Figure \ref{fig:Zygo}a presents cross-sectional height profiles taken along the line shown in Fig. \ref{fig:Zygo}b. The central horisontal lines represent the square silicon chip on which the metasurface is placed. Surrounding this planar region is some topography corresponding to the PZT membranes covered by electrodes. The mechanical shifting of these upon applying voltages is observed. Outside the membranes is the silicon chip to which the membranes are anchored. By alternating the application of 23V to the outer and inner electrodes, the central chip is displaced by 7.2$\mu$m vertically, while retaining a high degree of planarity throughout the movement. 

\section{Demonstration of a varifocal lens}
\label{sec:Varifocal}
\subsection{Metasurface design}
The metalenses consist of silicon rectangular pillars, as shown in Fig. \ref{fig:MEMS_device}b. The structures have lateral dimensions of 240nm $\times$ 330nm, height 1050nm and periodicity of 835nm.  The lens dimensions are $300\mu$m $\times 300\mu$m and are placed on a silicon substrate of thickness 500$\mu$m.

The geometrical phase principle is used for pointwise implementation of a lens phase function, working on circular polarization states of light \cite{kang2012wave, khorasaninejad2016metalenses}. The operation $T$ of the metalens on right ($|R\rangle$) and left ($|L\rangle$) circular polarized light, places the transmitted light in a superposition of circular polarization states according to

\begin{eqnarray}
T|R\rangle &=& B \exp(i2\alpha)|L\rangle + A|R\rangle, \label{eq:RPol} \\
T|L\rangle &=& A |L\rangle + B \exp(-i2\alpha)|R\rangle. \label{eq:LPol}
\end{eqnarray} For a suitable design of the metastructure one can achieve full cross-polarization ($|A|^2 \to 0$ and $|B|^2 \to 1$). The resulting cross-polarized field has attained a phase $2\alpha$ which can be shown to equal the rotation angle of the rectangular pillars (such rotations can be seen in Fig. \ref{fig:MEMS_device}b). The rotation $\alpha(r)$ is therefore varied pointwise over the radius $r$ of the metalens in order to impose the desired phase. The dimensions of the metasurface structure have been chosen to realize efficient cross-polarization for the target wavelength of $\lambda=1.55\mu$m.

\subsection{Varifocal setup}
\begin{figure}[ht!]
\centering\includegraphics[width=1.0\linewidth]{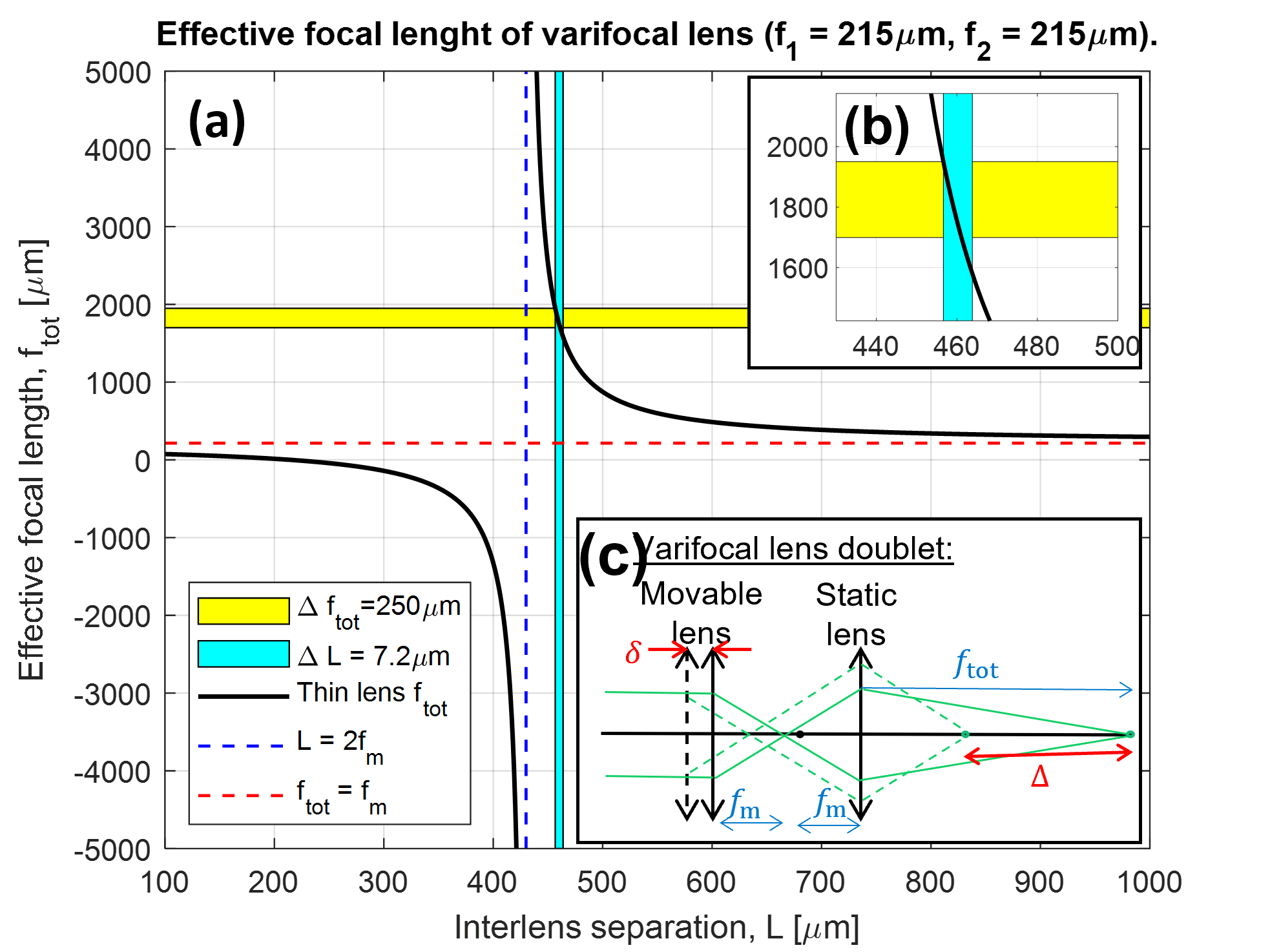}
\caption{(a) Effective focal length (EFL) vs. the interlens separation of two thin lenses with identical focal lengths under thin-lens paraxial assumptions. The EFL is defined as the optical path from the second lens to the focus point, as illustrated in (c). Dashed lines represent the asymptotes of the curve, which are related to the focal lengths of the individual lenses $f_m$. The yellow band represents the measured shift in EFL from the experimental demonstration (Sec. \ref{sec:results}). The blue band represents the measured displacement range $7.2\mu$m arbitrarily placed to intersect with the yellow band and $f_\text{tot}$, since we have not measured the interlens separation independently. (b) Magnified plot: it is observed that the MEMS displacement is slightly larger than that anticipated from the simple model. (c) Sketch of the varifocal setup: Modulating the interlens spacing by $\delta$ leads to a shift of the EFL by $\Delta$.}
\label{fig:EFLCurve}
\end{figure}

Figure \ref{fig:EFLCurve}c displays a simple thin-lens model for two lenses with identical individual focal lengths $f_m$ separated by an interlens separation of $L=L_0 + \delta$, where $\delta$ represents the displacement modulation which can be achieved by MEMS actuation and $L_0$ is separation when no actuation is applied. Incoming collimated light through the first lens is focused at the focal point $f_m$, i.e. a distance $L-f_m$ away from the second lens. Using the thin-lens equation (paraxial assumption), one finds that after passing the second lens the light is then focused to an effective focal point a distance

\begin{equation}
    f_\text{tot} = \frac{f_m(L-f_m)}{L-2f_m}, \label{eq:thinLens}
\end{equation} away from the second lens. Figure \ref{fig:EFLCurve}a plots the effective focal length (EFL) $f_\text{tot}$ against interlens separation $L$ assuming that the focal distances of the metalenses are $f_m=215\mu$m. Choosing a nominal distance $L_0$ close to the asymptote $L=2f_m$ ensures a large change in EFL ($\Delta$ in Fig. \ref{fig:EFLCurve}c) upon MEMS displacement $\delta$.


In order to realize this lens doublet, two identical metasurfaces are placed facing each other with their silicon substrates directed away from each other, as shown in Fig. \ref{fig:FocalShift}d. The metasurface to the left is MEMS displaced, whereas the other is statically held in place at a nominal distance $L_0$ from the first. The incident light on the first metasurface is right circularly polarized $|R\rangle$. According to \eqref{eq:RPol}, by transmission through the first metalens the pointwise phase $2\alpha_1(r)$ is applied to the cross-polarized light yielding: $\exp(i2\alpha_1(r))|L\rangle$ (assuming the metasurfaces are efficient, i.e. $|B|^2\to 1$). Consecutive transmission through the second lens then according to \eqref{eq:LPol} similarly leads to phase application to the cross-polarized counterpart yielding $\exp(i2[\alpha_1(r)-\alpha_2(r))|L\rangle$. Since the second metalens is identical but flipped relative to the first, the phase functions are equal apart from a sign change $\alpha_2(r)=-\alpha_1(r)$. Upon transmission through both metalenses, the twice cross-polarized light has therefore experienced the same phase application $\alpha_1(r)$ twice. The lens doublet described above and in Fig. \ref{fig:EFLCurve} is therefore implemented when $\alpha_1(r)$ is chosen to represent the phase function of a positive lens of focal length $f_m=215\mu$m.

\subsection{Results}\label{sec:results}
\begin{figure}[ht!]
\centering\includegraphics[width=1.0\linewidth]{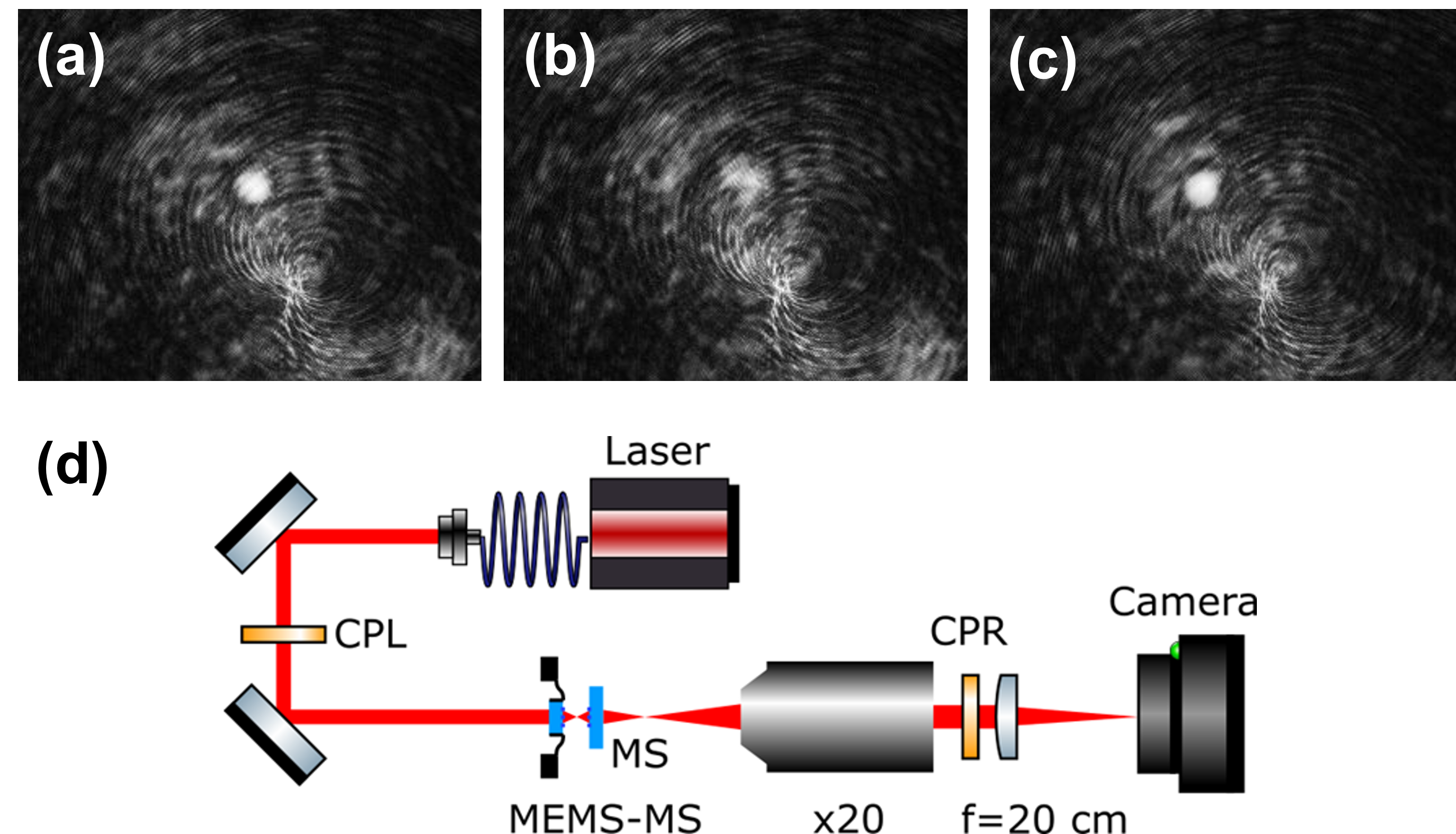}
\caption{Tuning the effective focal length of the metasurface lens doublet by MEMS actuation. (a) Imaging focal point when the MEMS metasurface lens is displaced to its highest position through applying 23V to \emph{outer} electrode of the MEMS chip. (b) Defocusing occurs upon displacing the metasurface lens to its lowest position through applying 23V to \emph{inner} electrodes. (c) Focus is regained by moving the objective lens 250$\mu$m away from lens doublet. This indicates that the focal point shifts by the same amount upon MEMS actuation. (d) Schematic of the optical setup}
\label{fig:FocalShift}
\end{figure}

To measure the focal tuning of the metalens doublet a setup as that sketched in Fig. \ref{fig:FocalShift}d is used. A 1.55$\mu$m fiber-laser is collimated and then reflected twice on tiltable mirrors with a left circular polarizing filter (CPL) in between them. This places the light in a right circular polarization state before passing through the metasurface doublet (the second mirror \emph{mirrors} the left circular polarization state transmitted from the CPL). The first metalens (MEMS-MS) is MEMS actuated. The resulting focal point after the second metasurface (MS) is imaged using a x20 infinity corrected objective, a tube lens and a IR-camera. Between the objective and the tube lens is placed a right circular polarizing filter (CPR). The objective is attached to a translation stage which is moved by a micro-screw, allowing for fine movements.

Figure \ref{fig:FocalShift}a and b show the respective focusing and defocusing caused by displacing the first metalens by PZT MEMS actuation. At focus/defocus 23V are applied to the outer/inner electrodes of the PZT membrane. The focal shift of approximately 250$\mu$m is measured by defocusing the lens and then measuring the distance the objective must be moved to regain focus (using the micro-screw on the translation stage). This is almost 35 times the MEMS displacement of 7.2$\mu$m measured in Fig. \ref{fig:Zygo} (although this was measured on a separate MEMS-chip of same design).

The EFL is found by first imaging the focal point, and then measuring the distance the objective must be moved in order to image the second metasurface. This gives the focal lengths $f_\text{outer}=1.95$mm and $f_\text{inner}=1.70$mm, corresponding to the EFLs when the outer and inner electrodes of the MEMS are actuated, respectively. Note that these values correspond to the \emph{optical} path length, in which refraction through the silicon substrate is taken into account. This is the relevant quantity for comparing with the thin-lens model by \eqref{eq:thinLens}. The \emph{physical} path length from the second metasurface to the focus points can be found by adding a (paraxial) correction factor of $t(1-1/n_\text{Si})\approx 0.36$mm to $f_\text{outer}$ and $f_\text{inner}$, where $t=500\mu$m is the silicon substrate thickness and $n_{Si}\sim 3.5$ is the refractive index. The change in optical power, calculated from $f_\text{outer}$ and $f_\text{inner}$, is 75m$^{-1}$.

In Fig. \ref{fig:EFLCurve}a and b the measured focal shift is represented by the yellow horizontal band in the plot, and the measured PZT MEMS displacement range is represented by the blue vertical band. The latter is arbitrarily placed to intersect with the yellow band and the solid curve (the exact interspacing between the metalenses has not been measured directly). Given the simplicity of the thin-lens model, acceptable qualitative correspondence is observed.



\section{Fabrication}
\label{sec:Fabrication}

The moveable MEMS component is fabricated by micro-structuring of a SOI wafer with thin-film PZT in similar steps to those described in \cite{bakke2010novel} for the fabrication of a MEMS micro-mirror using the same architecture. For the MEMS component in this work, however, the central part (the mirror element) has been removed while leaving a thin ledge around the opening. This makes it possible to mount a metalens on a silicon ledge inside the moveable part, while at the same time achieving good alignment and planarity.

The metasurface is fabricated using UV-nanoimprint lithography (NIL) to first pattern a resist mask (Microresist mr-NIL210-200nm) consisting of rectangular bricks on a silicon wafer. Subsequent dry etching was performed in a Rapier Si DRIE process module by SPTS, Newport, UK. The residual layer of the imprinted resist was first removed by ion bombardment in a continuous, directional etch with a pure Ar plasma. Finally a Bosch deep reactive ion etching (DRIE) step is used to etch the silicon rectangular pillars down to a depth of around 1µm. For more information on UV-NIL and Bosch DRIE consider \cite{dirdal2020towards, baracu2021metasurface}. In order to insert a metasurface lens into the ring of the MEMS actuator, the metasurfaces are diced into chips of 2.3mm x 2.3mm and carefully glued into place on the ledge. This manual and tedious process of fastening the metasurface to the MEMS actuator can be avoided by subsequent development for integrating the NIL patterning process into the MEMS fabrication process.

The MEMS chip with suspended metasurface chip is then glued to a PCB card and wirebonded. A controller card is then used to assign voltages to the six electrodes through the PCB.

\section{Discussion}
\label{sec:Discussion}
While the current MEMS chip offers large displacement relative to the current state of the art \cite{arbabi2018mems}, designs that offer significantly higher displacement are also possible. However there is a tradeoff between long displacement range and high switching speed.

The measured change in optical power in our setup is lower than that reported in the MEMS metasurface varifocal doublet demonstrations in \cite{arbabi2018mems}: While we have measured 75m$^{-1}$, an optical power change of more than 180m$^{-1}$ has been reported in \cite{arbabi2018mems}. According to the simple model in \eqref{eq:thinLens} shown in Fig. \ref{fig:EFLCurve}a, a larger change in optical focusing power than measured may be expected in our setup for an interlens spacing close to the vertical asymptote: On the order of 151m$^{-1}$ may be achieved when the interlens spacing is modulated within $L\in [430, 437.2]$. This is however still less than previously reported. An important reason for the lower optical power change in our measurement is the design choice of the focal length of the metasurfaces used here. In \cite{arbabi2018mems} the individual metalenses have focal lengths $120\mu$m and $-130\mu$m, while our metalenses have focal lengths of around $215\mu$m. \eqref{eq:thinLens} predicts that if we had adopted $f_m=120\mu$m in our setup, we could expect a change in optical power on the order of 472m$^{-1}$ for interlens modulation between $L\in[240\mu \text{m}, 247.2\mu\text{m}]$. The main reason that this predicted value is larger than that reported in \cite{arbabi2018mems} is due to the fact that our MEMS component achieves roughly twice the out of plane displacement.


\section{Conclusion}
\label{sec:Conclusion}
We have demonstrated first time use of integrated thin-film piezoelectric PZT for displacement of a dielectric metalens. Using PZT actuation allows for out of plane displacements of $7.2\mu$m with a voltage application of 23V. This is roughly twice the displacement at a quarter of the voltage application used in the state of the art electrostatic MEMS varifocal metalens. Our MEMS-metasurface component has been used to demonstrate a proof-of-concept varifocal setup with effective focal length tunability of around $250\mu$m.

\bibliographystyle{unsrt}  
\bibliography{Bibliography.bib}  





\end{document}